\documentclass[aps,pr,twocolumn,groupedaddress,notitlepage,
floatfix,superscriptaddress]{revtex4-2}

\pdfoutput=1 
\usepackage{graphicx,graphics,epsfig,subfigure,times,bm,bbm,amssymb,amsmath,amsthm,mathrsfs,MnSymbol} \usepackage{gensymb} \usepackage{amsfonts} \usepackage{float} \usepackage[matrix,frame,arrow]{xypic} \usepackage[pdfstartview=FitH]{hyperref} 
\usepackage{times}
\usepackage{float}
\usepackage{graphics}
\usepackage[T1]{fontenc}

\usepackage{braket}  
\usepackage{enumerate} \usepackage[normalem]{ulem}
\usepackage[usenames,dvipsnames]{xcolor} \usepackage{multirow} \usepackage{mathtools}
\usepackage{bbm} \usepackage{titletoc} 

\definecolor{orange}{rgb}{1,0.5,0} 
\hypersetup{ colorlinks=true,       
	linkcolor=red,          
	citecolor=blue,        
	filecolor=magenta,      
	urlcolor=blue,           
	runcolor=cyan }

\begin{document}
	\title{Effective Field Theories and Finite-temperature Properties of Zero-dimensional Superradiant Quantum Phase Transitions}
	
\author{Zi-Yong~Ge}
\affiliation{Theoretical Quantum Physics Laboratory, Cluster for Pioneering Research, RIKEN, Wako-shi, Saitama 351-0198, Japan}

\author{Heng~Fan}
\affiliation{Beijing National Laboratory for Condensed Matter Physics, Institute of Physics, Chinese Academy of Sciences, Beijing 100190, China}	
\affiliation{Songshan Lake Materials Laboratory, Dongguan 523808, Guangdong, China}
\affiliation{CAS Center for Excellence in Topological Quantum Computation, UCAS, Beijing 100190, China}
\affiliation{Beijing Academy of Quantum Information Sciences, Beijing 100193, China}

\author{Franco Nori}
\affiliation{Theoretical Quantum Physics Laboratory, Cluster for Pioneering Research, RIKEN, Wako-shi, Saitama 351-0198, Japan}
\affiliation{Quantum Information Physics Theory Research Team, Center for Quantum Computing, RIKEN, Wako-shi, Saitama 351-0198, Japan}
\affiliation{Department of Physics, University of Michigan, Ann Arbor, Michigan 48109-1040, USA}

\begin{abstract}
The existence of zero-dimensional superradiant  quantum phase transitions seems inconsistent with  conventional statistical physics.
This work clarifies this apparent inconsistency.
We demonstrate the corresponding effective field theories and finite-temperature properties of light-matter interacting systems,
and show how this zero-dimensional quantum phase transition occurs.
We first focus on the Rabi model, which is a minimum model that hosts a superradiant quantum phase transition.
With the path integral method, we derive the imaginary-time action of the photon degrees of freedom.
We also define a dynamical critical exponent as the rescaling between the temperature and the photon frequency, and perform dimensional analysis to the effective action.
The dynamical critical exponent shows that the effective theory of the Rabi model is a free scalar field,
where a true second-order quantum phase transition emerges.
These results are also verified by numerical simulations of imaginary-time correlation functions of the order parameter.
Furthermore, we also generalize this method to the Dicke model.
Our results make the zero-dimensional superradiant quantum phase transition compatible with conventional statistical physics,
and pave the way to understand it in the perspective of effective field theories.
\end{abstract}

	\maketitle
	\section{Introduction}
	
	Quantum phase transitions (QPTs) and quantum critical phenomena are two fundamental concepts in modern physics~\cite{stone2012physics,Fradkin2013,Sachdev2011,RevModPhys.69.315},
	and have been extensively studied in condensed matter physics~\cite{Sachdev2011,RevModPhys.69.315}, 
	high-energy physics~\cite{schwartzQFT}, and quantum information sciences~\cite{zeng2019quantum}.
	Recently, it is shown that there exists a novel  second-order QPT 
	from a normal phase to a superradiant phase in light-matter interacting systems~\cite{Scully1997,Agarwal2012}, such as the Dicke model~\cite{PhysRev.93.99,HEPP1973360,PhysRevA.7.831,PhysRevLett.90.044101,PhysRevLett.92.073602,
		PhysRevA.71.053804,PhysRevLett.112.173601,Wang_2014,PhysRevResearch.6.013303} and the Rabi model~\cite{PhysRev.49.324,PhysRevE.67.066203,PhysRevA.81.042311,PhysRevA.87.013826,PhysRevX.4.021046,
		PhysRevLett.115.180404,PhysRevLett.118.073001,PhysRevLett.119.220601,PhysRevA.100.063820,
		chen2021experimental,cai2021observation,arXiv:2207.12156,PhysRevLett.131.113601,PhysRevLett.132.053601}.
	The uniqueness of this superradiant QPT manifests in the ``dimension'' of these quantum optical systems, 
	where there is no spatial dimension.
	Meanwhile, special ``thermodynamic limits'' are also required for the true superradiant QPT.
	For instance,  the ``thermodynamic limit'' in the Rabi model is the photon frequency tending to zero~\cite{PhysRevLett.115.180404},  
	while it is the large atom limit in the Dicke model~\cite{HEPP1973360,PhysRevA.7.831}.
	The superradiant QPT can also be described by the language of spontaneous symmetry breaking, 
	where the corresponding  superradiant phase is a $\mathbb{Z}_2$ symmetry-breaking phase.
	Moreover, like conventional QPTs, the system can also host universal scaling laws near the superradiant critical point~\cite{PhysRevLett.119.220601},
	which is a strong evidence of a true second-order QPT.
	This superradiant QPT has attracted considerable interests 
	due to recent achievements of ultrastrong and even deep strong coupling regimes in light-matter interacting systems~\cite{PhysRevA.80.032109,niemczyk2010circuit,PhysRevLett.105.023601,PhysRevLett.105.263603,PhysRevLett.109.193602,PhysRevLett.112.016401,PhysRevA.94.033850,forn2017ultrastrong,Stefano_2017,yoshihara2017superconducting,PhysRevA.96.012325,PhysRevA.98.053819,di2019resolution,frisk2019ultrastrong,RevModPhys.91.025005,PhysRevResearch.3.023079,PhysRevA.103.053703,PhysRevLett.129.273602,PhysRevResearch.4.023048,PhysRevA.108.043717,10.21468/SciPostPhysLectNotes.50,Pilar2020thermodynamicsof}.

	In statistical physics, thermodynamic phase transitions are forbidden
	in 1D classical systems (or QPTs in 0D quantum systems), due to strong thermal (or quantum) fluctuations~\cite{Sachdev2011,Altland2010}.
	For instance, there is no ordered phase in the 1D classical Ising model at finite temperatures.
	From this viewpoint, the existence of the superradiant QPT in 0D quantum systems 
	seems inconsistent with conventional statistical physics.
	The general practice to understand the superradiant QPT is using perturbation theories to obtain a Gaussian-type effective Hamiltonian~\cite{PhysRevLett.115.180404},
	and the superradiant QPT can be identified by the spectrum of this solvable effective Hamiltonian.
	This perturbation theory is equivalent to the mean-field approximation, where the high-order terms correspond to the quantum fluctuations.
	According to statistical physics, the fluctuations in 0D quantum systems (or 1D classical systems) cannot be neglected.
	Therefore, it is still unclear why the mean-field approximation is valid in these 0D quantum systems,
	and this is significant for understanding the superradiant QPT~\cite{Sachdev2011,Altland2010}.
	The conventional Hamiltonian-based methods seem insufficient to explain why superradiant QPTs can occur.
	A natural question is whether we can use quantum field theories to describe the superradiant QPT, 
	making it more compatible with conventional statistical physics.
	In addition, finite-temperature physics is also a significant aspect to understand QPTs~\cite{Sachdev2011}.
	Hence, another open question is how superradiant QPTs affect finite-temperature behaviors in light-matter interacting systems.

	In this paper, we demonstrate how superradiant QPTs occur in 0D quantum-optical systems based on effective field theories.
	We first investigate the Rabi model, which is a minimum model to host a superradiant QPT.
	The imaginary-time action of the photon degrees of freedom is obtained by the path-integral method.
	In addition, we define a \textit{dynamical critical exponent} as the rescaling between the temperature and the photon frequency.
	By dimensional analysis, we find that, for a small dynamical critical exponent, the system is described by a massless free scalar field, 
	and is critical for an arbitrary coupling strength.
	For a marginal dynamical critical exponent,
     the effective theory  becomes a free scalar field or $\phi^4$-theory with mass term,
	and a true second-order QPT indeed emerges.
	For a large dynamical critical exponent, higher-order interaction terms become relevant, and the true QPT is absent.
	We also perform numerical simulations of imaginary-time correlation functions of the order parameter.
	Our numerical results show that the correlation function  increases when increasing the imaginary-time distance,
	which is a unique property of a 0D quantum phase.
	Meanwhile, the scaling of correlation functions is consistent with the effective field theory.
	By identifying the dynamical critical exponent with the gap at the critical point, 
	we can map this imaginary-time action to the ground state.
	We show that the effective theory is a free scalar field with a marginal 4th-order term.
	Thus, the mean-field approximation is indeed valid to describe the superradiant QPT in the Rabi model.
	Moreover, we also generalize this quantum-field method to the Dicke model.

    The remainder of the paper is organized as follows. In Sec.~\ref{sec2}, 
    we take the Rabi model as an example to show how a superradiant QPT occurs in a 0D quantum-optical system.
    We first review previous results of the superradiant QPT in the Rabi model, 
    then we derive the effective theory and use dimensional analysis to show how the superradiant QPT occurs.
    We also perform numerical simulations to support these analytical results.
    In Sec.~\ref{sec3}, we generalize this quantum-field method to the Dicke model.
    We present a discussion of our main results in Sec.~\ref{sec4}.
    We conclude in Sec.~\ref{sec5}.
    In the Appendix, we present more derivation details.

	\section{Rabi model}\label{sec2}
	
	Here we first consider the Rabi model, which is a minimum model to describe light-matter interacting systems~\cite{PhysRev.49.324}.
    Meanwhile, the Rabi model is also a minimum model to host a 0D superradiant QPT.
	In this section, we take the Rabi model as an example to show how the superradiant QPT occurs in a 0D quantum system.
   We mainly apply quantum field theories, and use numerical simulations to support our analytical results.
	
	\subsection{Review of the Rabi model}
	The Hamiltonian reads
	\begin{align} 
		\hat H_{\text{Rabi}}   = \omega \hat a^\dag \hat a + {\Omega}\hat \sigma^z+g \hat \sigma^x(\hat a^\dagger+\hat a),
	\end{align}	
   where $\hat a^\dagger$ ($\hat a$) is the creation (annihilation) operator of photons,
   $\hat \sigma^\alpha (\alpha=x,y,z)$ are the Pauli matrices describing the  two-level atom,
   $\omega$ is the frequency of the photon, $\Omega$ is the gap of the atom, and $g$ is the strength of the dipole interaction.
   The Hamiltonian $\hat H$ possesses parity symmetry
   	\begin{align} 
   	[\hat P, \hat H_{\text{Rabi}} ]=0, \ \  \hat P = (-1)^{\hat n}\hat \sigma^z,
   \end{align}	
  where $\hat n = \hat a^\dagger\hat a$ is the number operator.
   In addition, we can define a dimensionless parameter 
   	\begin{align} 
    \lambda:=\frac{2g}{\sqrt{2\omega \Omega}},
   \end{align}	
    which describes the characteristic light-matter coupling strength.

   Previous works~\cite{PhysRevLett.115.180404} show that, under the special thermodynamic limit 
   	\begin{align} 
   	\frac{\Omega}{\omega_0}\rightarrow 0,
   \end{align}
   there exists a second-order QPT in this 0D quantum system when varying $\lambda$.
   The critical point is exact at $g_c=1$, and this QPT can be described by the spontaneous breaking of the parity symmetry.
   In the regime $g<g_c$, the system is in its normal phase with no symmetry breaking.	
   However, when $g>g_c$, the system is in the superradiant phase, 
   where the parity symmetry is spontaneously broken.
   Moreover, like conventional critical phenomena, this system also hosts universal scaling laws near the critical point~\cite{PhysRevLett.119.220601}.
   Hereafter, we derive the effective field theory of the Rabi model to show how this 0D QPT occurs.

   	\begin{figure*}[t] \includegraphics[width=1\textwidth]{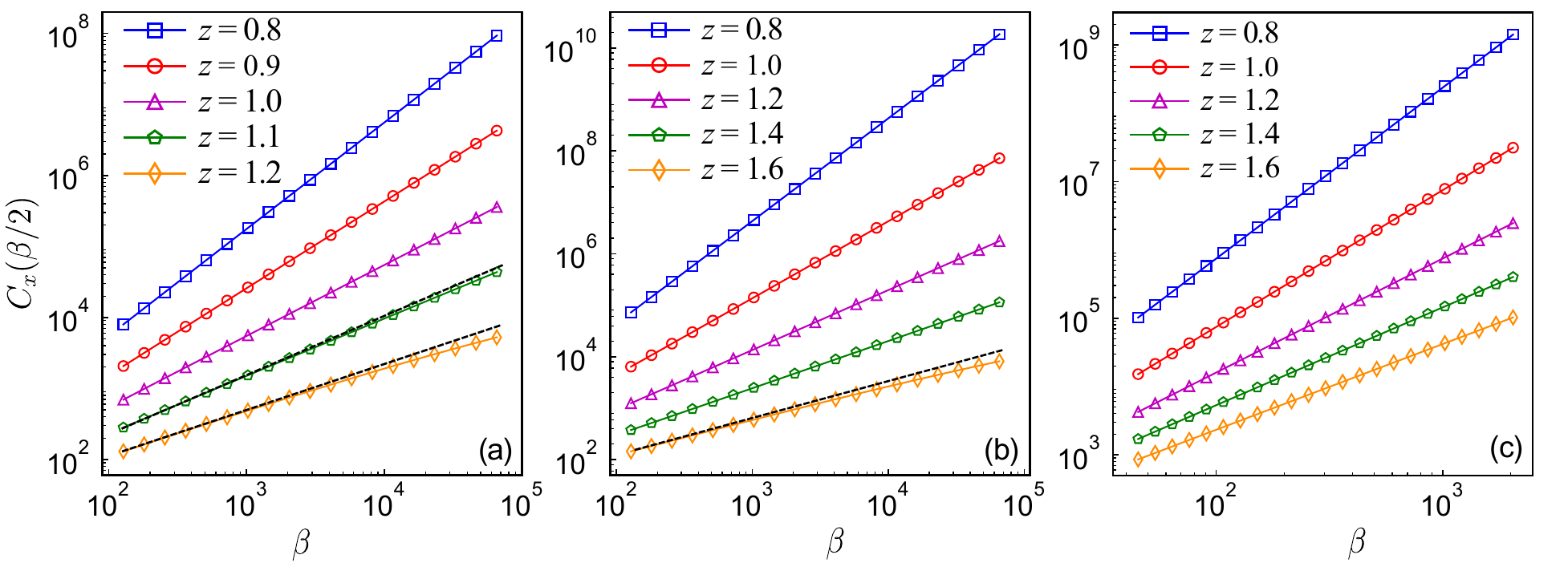}
   	\caption{Imaginary-time correlation function of displacement operator $C_x(\beta/2)$.
   		(a) $\lambda=0.8$. When $z'\leq1$, $C_x(\beta/2)$ shows a power-law increase.
   		Two black dashed lines are linear fits. 
   		(b) $\lambda=\lambda_c=1$. In this case, $C_x(\beta/2)$ has a power-law increase when $z'\leq1.5$.
   		The black dashed line is a linear fit.
   		(c) $\lambda=1.2$. Here, $C_x(\beta/2)$ has a power-law increase for all cases.}
   	\label{fig_1}
   \end{figure*}

   \subsection{The effective action}
	For simplicity, we use a harmonic oscillator to describe the photon degrees of freedom, and represent the spin operator by a spinor~\cite{popov1988functional,PhysRevB.69.113203}.
	Hence, the Hamiltonian can be rewritten as (choosing oscillator mass $m=1$)
	\begin{align} \label{Hrabi2}
		\hat H_{\text{Rabi}}   = \frac{1}{2}\hat p ^2 + \frac{1}{2}\omega^2 \hat x^2 + \Omega\hat \psi^\dagger \sigma^z\hat\psi + \frac{1}{2}\lambda\omega\sqrt{2\Omega}\hat x\hat \psi^\dagger \sigma^x\hat\psi,
	\end{align}	
	where $ \hat\psi^\dagger = [\hat c_\uparrow^\dagger,\hat c_\downarrow^\dagger]$ is a spinor with $\hat c_{\uparrow,\downarrow}^\dagger$
	being a spin-$\frac{1}{2}$ fermion creation operator,
	$\hat x = (\hat a^\dagger+\hat a)/\sqrt{\omega}$ is the displacement operator of the harmonic oscillator, 
	and $\hat p = i\sqrt{\omega}(\hat a^\dagger-\hat a)$ is the momentum operator.
	Here, we apply the path integral method to obtain the effective field theory of the photon~\cite{Altland2010}.
	The partition function of the system takes the form
	\begin{align} 
		Z = \int \!\!  \mathcal{D}x \; \mathcal{D}\bar \psi \;\mathcal{D}\psi \; \exp\big[-S(x, \bar \psi, \psi)\big].
	\end{align}	
	According to Eq.~(\ref{Hrabi2}), the imaginary-time action can be obtained as 
	\begin{align} 
		S(x, \bar \psi, \psi) =S_0(x) +S_1(x, \bar \psi, \psi),
	\end{align}	
   where $S_0$ is the free term, and $S_1$ is the interacting term, with the detailed forms 
   \begin{subequations}
	\begin{align} 
	&S_0(x)= \int_0^\beta d\tau  \bigg[ \frac{1}{2} (\partial_\tau x) ^2 +   \frac{1}{2}\omega^2   x^2  \bigg], \\ 
	&S_1(x, \bar \psi, \psi) = \int_0^\beta \! d\tau\;  \bar \psi (\partial_\tau + {\Omega} \sigma^z+ \frac{1}{2}\lambda\omega\sqrt{2\Omega} x  \sigma^x)\psi,
  \end{align}	
\end{subequations}
  where $\beta =1/T$ is the inverse of the temperature ($k_b=1$), 
  $x$ is a real scalar field representing the coordinate of the oscillator, and $ \bar \psi (\psi)$ is a spinor field.

 When $\beta\rightarrow\infty$, the action $S$ describes the ground-state physics of the Rabi model.
 Integrating out the spinor field, we can obtain the effective action of the oscillator as
	\begin{align} \label{Seff}\nonumber
		S_{\text{eff}} (x)= \int_0^\beta d\tau  \bigg[& \frac{1}{2} \big(1+ \frac{\lambda^2\omega^2}{4\Omega^2}\big)(\partial_\tau x) ^2 + 
		\frac{1}{2}(1-\lambda^2)  \omega^2x^2 \\
		&+ a_4 \omega^4 x^4+  a_6 \omega^6 x^6+.... \bigg],
	\end{align}	
	where the factor $a_{2n}$ is a function of $\Omega$ and $\lambda$, and the explicit form is unimportant.
	The detailed derivations of $S_{\text{eff}} (x)$ are presented in Appendix~\ref{app1}.
	There exists a parity symmetry in Eq.~(\ref{Seff}), i.e., $S_{\text{eff}}$ is invariant under the transformation $x \rightarrow -x$, 
	which corresponds to the original parity $\hat P$ in $\hat H_{\text{Rabi}}$.

    \subsection{Dimensional analysis and the phase transition}
    
	Now we perform the dimensional analysis~\cite{Altland2010} of the effective action $S_{\text{eff}}$.
	As usual, the kinetic term is set to be unity: $\big[\int d\tau  (\partial_\tau x) ^2\big]=1$,
   where $[F(x)]\sim \beta^{d_F}$ implies that the canonical dimension of the term $F(x)$ is $d_F$.
	Thus, the canonical dimension of the scalar field $x$ and other terms satisfy
	\begin{align} \label{Dx_canon}
	[x]=\beta^{1/2}, \ \ \ \big[\int d\tau x ^{2n}\big]=\beta^{n+1}.
	\end{align} 
	Therefore, if the frequency $\omega$ is finite (i.e., dimensionless), then an arbitrary order of the interaction term is relevant, 
	which cannot be neglected in the renormalization group (RG) flow.
	In this case, the mean-field theory and $\phi^4$-theory are both invalid,
	and there should not exist any true second-order QPT.
	This is also the reason why second-order QPTs are  generally absent in 0D quantum systems 
	(or thermodynamical phase transitions in 1D classical systems).

	However,  in the ``thermodynamic limit'' $\omega\rightarrow0$, a continues QPT indeed exists in the Rabi model at the critical point $\lambda_c=1$.
	Now we understand this special QPT from the viewpoint of the effective action in Eq.~(\ref{Seff}).
	A conventional $d$-dimensional ($d\geq1$) quantum critical  phenomenon can be described 
	by the $(d+1)$-dimensional imaginary-time action under the condition of $\beta\sim L^z$~\cite{Sachdev2011}, 
	where $L$ is the system size and $z$ is the \textit{dynamical critical exponent}.
	Meanwhile, the finite size in critical systems corresponds to an infrared (IR) cutoff $\xi \sim L\sim \beta^{1/z}$~\cite{Sachdev2011}, where $\xi$ is the maximal wavelength.
	Therefore, in the Rabi model, if we regard the finite photon frequency as an IR cutoff, 
	then considering the condition $\omega\rightarrow0$ as a ``thermodynamic limit'' is reasonable in  this picture.

	To describe the quantum critical phenomenon of the Rabi model by $S_{\text{eff}}$, we need to define an analogous ``dynamical critical exponent'' $z'$.
	Here, the maximal wavelength is $\xi\sim1/\omega$, so this dynamical critical exponent can be defined as the 
	rescaling between $\omega$ and $\beta$,
	\begin{align}
		\omega\sim \beta^{-1/z'}.
	\end{align} 
	Thus, $z'$  can also be understood as giving $\omega$ a non-zero dimension, i.e., $[\omega]=\beta^{-1/z'}$.
	Since $\beta\rightarrow\infty$ for the ground state, the photon frequency also satisfies $\omega\sim \beta^{-1/z'}\rightarrow\infty$,
	which recovers the thermodynamic limit of the Rabi model.
	The mass term and interaction terms  now have the dimension
	\begin{align} \label{dim}
		\big[\omega^{2n}\!\!\int \!\! d\tau x ^{2n}\big]=\beta^{n-2n/z'+1}.
	\end{align}	
  Therefore, these terms are now not necessarily relevant when the dynamical critical exponent $z'$ is small enough,
  and the effective theory can be truncated to finite order.
  Note that $z'$ is a constant in a specific model.
  In the following, we first consider $z'$ as a tunable parameter at finite temperatures, and discuss the corresponding effective theory.
  We will confirm $z'$ in the Rabi model when discussing ground state properties in Sec.~\ref{sec2}.~E.

  According to Eq.~(\ref{dim}), we can find that different $z'$ may lead to different effective theories of the Rabi model.
  When $0<z'<1$, i.e., $(n-2n/z'+1)<0$ for $n\geq 1$, the mass term and all interaction terms become irrelevant.
  In this case, the system is described by a massless free scalar field, leading to a critical phase for an arbitrary $\lambda$.
  
  When $1<z'<4/3$, the mass term is relevant, while all interaction terms are irrelevant.
  Thus, the effective theory is a free scalar field with relevant mass term, and the system hosts a critical point at $\lambda=1$, 
   where the mass of the scalar field vanishes.
   
   For  $4/3<z'<3/2$, the $\omega^{4}x^4$ term becomes relevant with higher-order interaction terms irrelevant, 
   so the effective action is a $\phi^4$-theory, which also hosts a critical point at $\lambda= 1$.

  When $z'>3/2$, higher-order interaction terms are also relevant, and the true second-order QPT is absent.
	
		\begin{figure}[t] \includegraphics[width=0.45\textwidth]{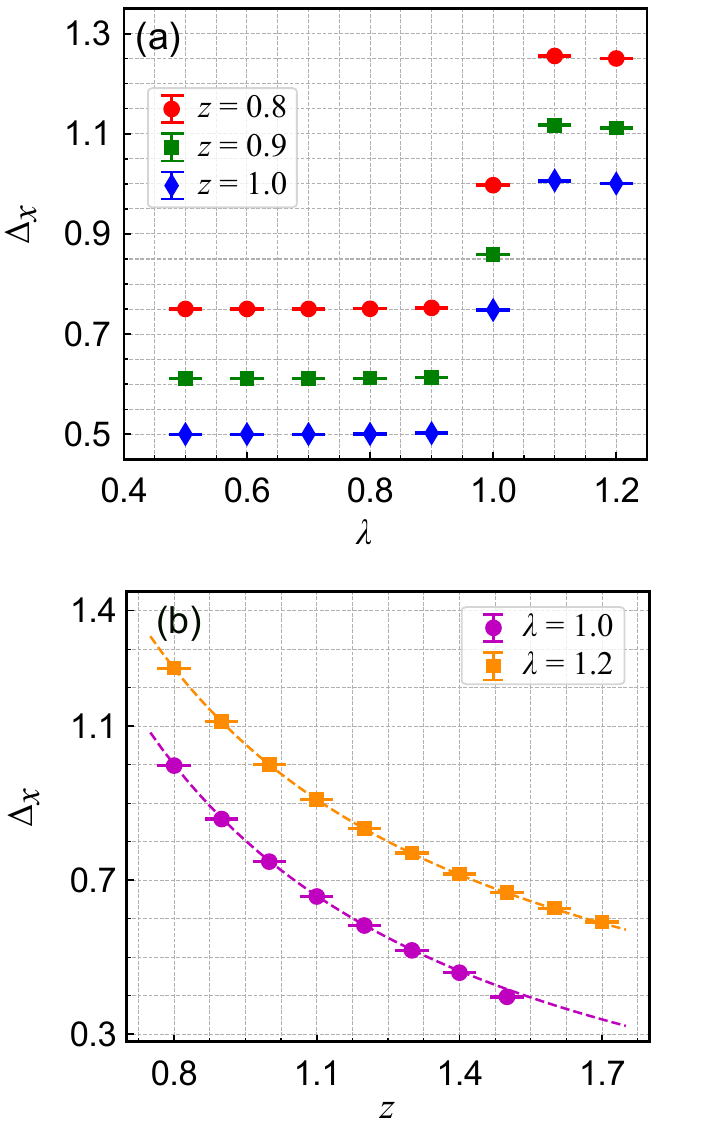}
		\caption{The observed dimension of the displacement operator.
			(a) $\Delta_x$ versus $\lambda$ for $z'\leq 1$. There is a sudden change for $\Delta_x$ at $\lambda=1$.
			When $\lambda<1$,  $\Delta_x\approx1/z'-1/2$; when  $\lambda=1$,  $\Delta_x\approx1/z'-1/4$; and  $\Delta_x\approx1/z'$ for $\lambda>1$.
			(b) $\Delta_x$ versus $z'$ for $\lambda\geq 1$. The orange dashed curve is the function $\Delta_x=1/z'$, while the magenta one is $\Delta_x=1/z'-1/4$.}
		\label{fig_2}
	\end{figure} 

  \subsection{Numerical simulations}
    We have presented an effective field theory and the dimensional analysis of the Rabi model  to show how superradiant QPTs occur in this 0D quantum system.
    To support the above discussions and further understand superradiant QPTs,
	 we perform numerical simulations of the finite-temperature system.
	Here we mainly calculate correlation functions of the order parameter $\hat{x}$ in the imaginary-time coordinate, defined as
		\begin{align} 
		C_{x}(\tau) : = \braket{\hat{x}(\tau)\hat{x}(0)}_\beta.
	\end{align}	
	Here, $\hat{\mathcal{O}}(\tau)= e^{\tau\hat H}\hat{\mathcal{O}}e^{-\tau\hat H}$ is the operator in the Heisenberg picture at imaginary time $\tau$, 
	and $ \braket{\cdot}_\beta:=\text{Tr}(\cdot e^{-\beta\hat H})/\text{Tr}( e^{-\beta\hat H})$ is the thermodynamic average at  temperature $1/\beta$.
	Meanwhile, we also fix the relation between the temperature and photon frequency as $\beta=\omega^{-z'}$.
	During the numerical calculation, the parity symmetry is preserved in the density matrix, so we have $\braket{\hat x(\tau)}_\beta=0$.
	Thus, in the ordered phase, $C_{x}(\tau\rightarrow\infty) = \bar x^2$, 
	where $ \bar x$ is the susceptibility, i.e., the expectation value of $\hat x$ in either parity symmetry broken state.
	If the system is critical, then the correlation function $C_{x}(\tau)$ exhibits a power-law decay/increase as $\tau$ increases due to scale invariance
	\begin{align} 
		C_{x}(\tau)\sim\tau^{2\Delta_{x}},
	\end{align}	
	 where $\Delta_x$ is the observed dimension of the operator $\hat{x}$.
	To reduce the finite-size (finite-$\beta$) effect, 
	we calculate the ``half-chain'' correlation function $C_{x}(\beta/2)$ for different $\beta$,
	and study the relation between $C_{x}(\beta/2)$ and $\beta$.
	In addition, we enlarge the Fock space of the photon until the result converges.

	The ``half-chain'' correlation function $C_{x}(\beta/2)$ versus $\beta$ for different $\lambda$ and $z'$ is presented in Fig.~\ref{fig_1}.
	We can find that $C_{x}(\beta/2)$ exhibits an increase when increasing $\beta$ for all cases, which is distinct to conventional high-dimensional systems,
	where the correlation function should decay for increasing distance.
    Equation~({\ref{Dx_canon}}) shows that the canonical dimension of the operator $\hat x$ is positive.
    Thus, generally, the correlation function $C_{x}(\beta/2)$ tends to increase when increasing the size $\beta$.
    However, for higher dimensional quantum systems ($d\geq1$), the corresponding order parameter has the canonical dimension $(1-d)/2\leq0$,
    so the correlation function tends to decay for increasing $\beta$.
	Therefore, this result, i.e., \textit{the correlation function of the order parameter increases as $\beta$ increases}, 
	is a unique property of 0D quantum phases.

	Now we discuss the correlation function $C_{x}(\beta/2)$ in detail.
	When $\lambda < 1$, Fig.~\ref{fig_1}(a) shows that $C_{x}(\beta/2)$ hosts a power-law increase only when $z'\leq1$, with the observed dimension $\Delta_{x}=1/z'-1/2$  [see Fig.~\ref{fig_2}(a)].
	This demonstrates that, when $z'\leq1$, the system is described by a massless scalar field even for $\lambda<1$,
	i.e., the mass term in Eq.~(\ref{Seff}) is indeed irrelevant  in this case.
	
	In the case of $\lambda = 1$,  $C_{x}(\beta/2)$ can exhibit a power-law increase when $z'\leq3/2$, see Fig~\ref{fig_1}(b).
	This is a strong evidence that there is no true second-order QPT at $\lambda=1$ when $z'>3/2$,
	i.e, the superradiant criticality is absent in this case.
	Moreover, we also fit the dimension of $\hat x$ at the critical point as $\Delta_{x}=1/z'-1/4$ [Fig.~\ref{fig_2}(b)].
	
	When $\lambda > 1$,  $C_{x}(\beta/2)$ shows a power-law increase for an arbitrary $z'$, see Fig.~\ref{fig_1}(c).
	In addition, the dimension of $\hat x$ is $\Delta_x = 1/z'$ in this case, see Fig.~\ref{fig_2}(b).

	The observed dimension of $\hat x$ versus $z'$ and $\lambda$ is summarized in Table~\ref{tab1}.
	Now we understand numerical results of $C_{x}(\beta/2)$ in terms of the dynamical critical exponent $z'$,
	where the system can be divided into three regions.
	When $0<z'\leq1$, the system is critical for an arbitrary $\lambda$, and there is no second-order QPT at $\lambda=1$, 
	though $\Delta_{x}$ is not continuous at $\lambda=1$.
	Here, the sudden change of $\Delta_{x}$ originates from the exact zero of the mass term at $\lambda=1$.
	For $1<z'\leq3/2$, there is a true second-order QPT at $\lambda=1$, 
	where the susceptibility in the ordered phase is $\bar x \sim \beta^{1/z'}\sim \omega^{-1}$.
	When $z'>3/2$,  the true superradiant QPT is absent, since there is no scale invariance at $\lambda=1$.
	Therefore, the numerical results are consistent with the effective theory and dimensional analysis.
	
	\begin{table}[t]
		\caption{\label{tab1}
			The observed dimension of the displacement operator $\Delta_x$. 
			Here, ``NSI'' means ``no scale invariance'' in the corresponding case.}
		\begin{ruledtabular}
			\begin{tabular}{cccccccc}
				&$0<z'\leq1$    &$1<z'\leq3/2$  &$z'>3/2$  \\ \hline
				$\lambda<1$ & $1/z'-1/2$      & NSI                  & NSI          \\
				$\lambda=1$ & $1/z'-1/4$    & $1/z'-1/4$    & NSI          \\
				$\lambda>1$ & $1/z'$              & $1/z'$              & $1/z'$      \\
			\end{tabular}
		\end{ruledtabular}
	\end{table}

   \subsection{Mapping to the ground state}
\begin{figure}[t] \includegraphics[width=0.45\textwidth]{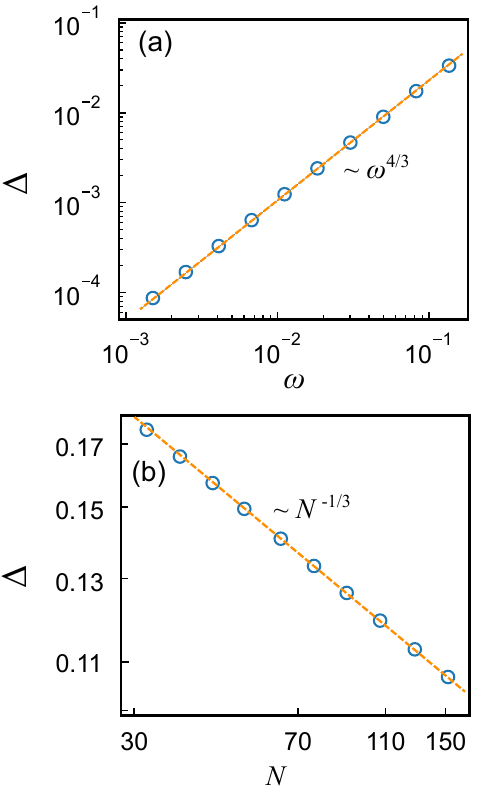}
	\caption{The scaling of the gaps in (a) the  Rabi model and (b) the Dicke model, respectively.
	The  orange dashed lines are linear fittings.}
	\label{fig_3}
\end{figure}

   We have studied the effective theory of the Rabi model at finite temperatures.
   Now we apply the above results to the ground state.
   Here, the key is to confirm the dynamical critical exponent $z'$ in the Rabi model.
   According to conventional QPTs, the definition of dynamical critical exponent 
   can also be expressed as the scaling between the gap and the length scale, i.e., $\Delta\sim\xi^{-z}$.
   Similarly, in the Rabi model, the length scale at the critical point is $\xi\sim1/\omega$,
   and $z'$ can be confirmed by the relation
   	\begin{align} 
   	\Delta\sim\omega^{z'}.
    \end{align}	
	Now we numerically obtain the dynamical critical exponent  as
	\begin{align} 
		z'=4/3,
	\end{align}	
	which is consistent with Ref.~\cite{PhysRevLett.115.180404}, see also our Fig.~\ref{fig_3}(a).
	
	According to Eq.~(\ref{dim}), when $z'=4/3$, the mass term is relevant, the $\omega^4x^4$ term is marginal, and other higher-order terms are irrelevant.
	Thus, the system can be described by a 1D real free scalar field.
	Since the fluctuation terms are all negligible, the mean-field approximation is indeed valid to understand the superradiant QPT.
	
	To further uncover the phase transition by the effective action, we can apply the principle of least action.
	For simplicity, we use the following simplified effective action of the free scalar field
	\begin{align} 
		S_{\text{eff}}= \int_0^\beta d\tau  \bigg[\frac{1}{2} (\partial_\tau x) ^2 + 
		\frac{1}{2}(1-\lambda^2)  \omega^2x^2 \bigg].
	\end{align}	

	(\textit{i}) For positive mass, i.e., $|\lambda|<1$, to minimize the action we have $x=0$.
	Thus, in this case, the parity symmetry is unbroken, corresponding a normal phase.
	
	(\textit{ii}) For negative mass, i.e., $|\lambda|>1$, to minimize the action  we have $x=\infty$.
	This shows that there exists an instability leading to a condensation of $x$, which corresponds to a symmetry-breaking phase.
	When mapping to the original Hamiltonian, we can know this symmetry-breaking phase is indeed a superradiant phase.
	
	We can also obtain the scaling of the order parameter $\hat x$ at the critical point
	\begin{align} 
		\braket{\hat x}\;\sim\; \beta^{1/2}\;\sim \; \omega^{-2/3},
	\end{align}	
   which is consistent with the result in Ref.~\cite{PhysRevLett.115.180404}.

	\section{Dicke model}\label{sec3}
	
	Now we consider the Dicke model, which describes light interacting with a large ensemble of two-level atoms~\cite{PhysRev.93.99}.
	The Hamiltonian reads
	\begin{align} 
		\hat H_{\text{Dicke}} = \omega \hat a^\dag \hat a + {\Omega}\sum_{j=1}^N\hat \sigma^z_j+\frac{2J}{\sqrt{N}}(\hat a^\dagger+\hat a)\sum_{j=1}^N\hat \sigma^x_j.
	\end{align}	
	where $\hat \sigma^\alpha_j (\alpha=x,y,z)$ describe the $j$-th two-level atom,
	and $N$ is the number of atoms.
	Parity symmetry implies 
	\begin{align} 
		[\hat P, \hat H_{\text{Dicke}}]=0, \ \ \ \hat P = (-1)^{\hat n}\prod_{j=1}^N\hat \sigma_j^z.
	\end{align}	
    In the thermodynamic limit $N\rightarrow \infty$, 
    there also exists a second-order QPT in this 0D system from the normal phase to the superradiant phase when increasing $\lambda$~\cite{HEPP1973360,PhysRevA.7.831}.
   The critical point is exact at 
   	\begin{align} 
   \lambda=\lambda_c=\sqrt{\omega\Omega}/2,
   \end{align}	
   and this QPT can also be described by the spontaneous breaking of the parity symmetry.

	Now we apply the above method in the Rabi model to discuss how the superradiant QPT occurs in the Dicke model.
	Here, the effective imaginary-time action can be obtained as
	\begin{align} \label{Seff2}\nonumber
		S_{\text{eff}} (x)= \int_0^\beta d\tau  \bigg[& \frac{1}{2} (\partial_\tau \tilde{x}) ^2 + 
		\frac{2\omega}{\Omega}(\lambda_c^2-\lambda^2)  \tilde{x}^2 \\
		&+ \frac{\alpha_4}{N}  \tilde{x}^4+   \frac{\alpha_6}{N^2} \tilde{x}^6+.... \bigg],
	\end{align}	
	where $\tilde{x}$ is also a real scalar field, and  $\alpha_{2n}$ is a finite factor with the explicit form unimportant.
	The detailed derivations are presented in Appendix~\ref{app2}.
	Similar to the Rabi model, the canonical dimension of the field $x$ is also $[\tilde{x}]=\beta^{1/2}$.
	Since $\omega$ and $\Omega$ are both finite in the Dicke model, the mass term is always relevant, i.e., $[\int d\tau \tilde{x}^2]=\beta^{2}$.
	In the case of finite size, i.e., finite $N$, each interaction term is relevant, so there should be no true second-order QPT.

	In the ``thermodynamic limit'' $N\rightarrow\infty$, analogous to the Rabi model, we can define a dynamical critical exponent as $N\sim\beta^{1/z'}$.
	Thus the dimension of the interaction terms are now
		\begin{align} 
		\big[N^{n-1} \!\!\int \!\! d\tau \tilde{x} ^{2n}\big]=\beta^{n+1-(n-1)/z'}.
	\end{align}	
   When $z'<1/3$, all interaction terms become irrelevant, and the system can be effectively described by the free real scalar field with a mass term.
   
   For $1/3<z'<1/2$, the $x^4$ term is relevant, while higher-order interaction terms are irrelevant,
   and the effective theory is a $\phi^4$-theory.
   In the above two cases, there exists a QPT at the critical point $\lambda=\lambda_c$.
   
   However, when $z'>1/2$, the higher-order interaction terms are relevant, leading to the absence of a true QPT.
   Therefore, the superradiant QPT of the Dicke model can also be described by the effective field theory in Eq.~(\ref{Seff2}).

  Here, we discuss the ground-state properties of the Dicke model, where we need to know the dynamical critical exponent $z'$.
  Similar to the Rabi model, $z'$ can be calculated by the relation 
  	\begin{align} 
  	\Delta\sim N^{-z'},
  \end{align}	
  where $\Delta$ is the gap.
  Via numerical simulations, we obtain $z'=1/3$, see Fig.~\ref{fig_3}(b).
  Thus, the $\tilde{x}^4$ term is marginal, and other higher-order terms are irrelevant.
  Therefore, similar to the Rabi model, the effective theory of the Dicke model can be described by the mean-field approximation, 
  and there indeed exists a QPT when tuning the mass from positive to negative.

   \section{Discussion}\label{sec4}
   
   The correlation length is divergent at the critical point, 
   so the long-wavelength physics is a significant aspect to understand QPTs~\cite{Sachdev2011}.
   However, the original Hamiltonian generally contains too many microscopic details, which are not useful for analyzing long-wavelength physics.
   Thus, to further understand the QPT, we can integrate out all of the microscopic details to obtain an effective field theory.
   This long-wavelength-limit effective field theory can be used to provide insight on the QPT.
   Specifically, it can tell us which symmetry dominates the QPT, and how the quantum fluctuations impact the phases.
   This is common practice in many conventional QPTs.
   Here, to understand how superradiant QPTs occur in 0D systems, we also introduce effective field theories in the long-wavelength limit.

   According to Eq.~(\ref{Seff}), we can find that the high-order term $\omega^{2n}x^{2n}$ (the quantum fluctuations) contains a prefactor $\omega^{2n}$.
   In the thermodynamic limit $\omega\rightarrow0$, it seems that $\omega^{2n}x^{2n}$ also tends to zero,
   i.e., the mean-field approximation is convergent.
   However, for this 0D system, the expectation value of the field $x$ is divergent, i.e., $x^{2n}\rightarrow\infty$, see Fig.~\ref{fig_1}.
   Thus, the fluctuation term $\omega^{2n}x^{2n}$ is not necessarily convergent.
   From this viewpoint, the validity of the mean-field approximation for the Rabi model is not obvious.
   To address this puzzle, we study its effective field theory to analyze whether the quantum fluctuations are relevant.
   Here, the core idea is introducing the dynamical critical exponent $z'$ to relate two divergent quantities, i.e., $1/\omega$ and $x^{2n}$.
   By dimensional analysis, we find that the $\omega^{4}x^{4}$ term is marginal, while the higher-order terms are irrelevant.
   Therefore, the quantum fluctuation is indeed negligible, and the mean-field approximation is valid to describe this 0D superradiant QPT.

	\section{Summary}\label{sec5}
	
	In conclusion, we have investigated the effective theories and finite-temperature properties of the superradiant QPT, 
	and shown how these occur in 0D light-matter interacting systems.
	Using the path integral method, we first derive the effective imaginary-time action of the photon in the Rabi model.
	We also define the dynamical critical exponent as the rescaling between the temperature and the photon frequency.
	We perform a dimensional analysis to discuss whether high-order terms are relevant, and the results show:
		
	(\textit{i}) When the dynamical critical exponent is small enough, the system is described by a massless free scalar field, 
	leading to a critical phase for an arbitrary coupling strength.
	
	(\textit{ii}) The effective theory can be a free scalar field or $\phi^4$-theory with mass term for a marginal dynamical critical exponent,
	and a true second-order QPT indeed emerges.
	
	(\textit{iii}) For a large dynamical critical exponent, higher-order interaction terms become relevant, and the true QPT is absent.
	These results were also verified by numerical simulations.
	
	We also numerically obtain  the dynamical critical exponent, 
	and our results show that the superradiant QPT in the Rabi model can indeed be described by a mean-field approximation.
	We also generalize this quantum-field method to the Dicke model.

	Our results make the 0D superradiant QPT compatible with conventional statistical physics,
	and pave the way to understand the superradiant criticality from the view point of effective field theories.
	There also remain several interesting topics that deserve further study:
	We find that the 4th-order terms are marginal for both the Rabi model and Dicke model,
	and it would be an interesting issue to explore whether it is universal in 0D or systems with all-to-all connectivity.
    In addition, it will also be relevant to obtain the universal scaling law of 0D superradiant QPTs by effective field theories.
	Other meaningful topics include whether our field-theory methods can be generalized to dispersive QPTs in light-matter interacting systems~\cite{PhysRevLett.51.1506,SCHON1990237,PhysRevA.97.013825,PhysRevLett.130.210404}, 
	and what will be happen to the effective theory when adding the spatial dimension, i.e., considering the Rabi lattice~\cite{PhysRevLett.99.186401,PhysRevLett.109.053601,PhysRevLett.109.053601}.

	\begin{acknowledgements}
		F. N. acknowledges support from:
		Nippon Telegraph and Telephone Corporation (NTT) Research,
		the Japan Science and Technology Agency (JST) [via
		the Quantum Leap Flagship Program (Q-LEAP), and
		the Moonshot R\&D Grant Number JPMJMS2061],
		the Asian Office of Aerospace Research and Development (AOARD) (via Grant No. FA2386-20-1-4069), and
		the Office of Naval Research (ONR).
		 H. F. acknowledges support from the National Natural Science Foundation of China (via Grant No. 92265207 and No. T2121001).
	\end{acknowledgements}

	\begin{appendix}
		
	\section{Effective action of the Rabi model}\label{app1}
	
	Here we derive the effective action of the Rabi model.
	For simplicity, we rewrite the effective Hamiltonian in terms of the harmonic  oscillator
	\begin{align} 
		\hat H  = \frac{1}{2}\hat p ^2 + \frac{1}{2}\omega^2 \hat x^2 + \Omega\hat \sigma^z+g\sqrt{\omega}\hat x\hat \sigma^x.
	\end{align}	
	We can also define the spinor $ \hat\psi^\dagger = [\hat c_\uparrow^\dagger,\hat c_\downarrow^\dagger]$, where $\hat c_\alpha^\dagger$ is the fermion operator.
	Thus, the spin operator can be obtained as
	\begin{align} 
		\hat \sigma_{\alpha\gamma} = \hat\psi^\dagger \sigma_{\alpha\gamma}  \hat\psi.
	\end{align}	
	Expressed in the form of path integral, the partition function of the system takes the form
	\begin{align} 
		Z = \int Dx \; D\bar \psi \; D\psi \;\exp\big[{-S(x, \bar \psi, \psi)}\big],
	\end{align}	
	where the action $S(x, \bar \psi, \psi)$ has the form
	\begin{subequations}
	\begin{align} 
		&S(x, \bar \psi, \psi) =S_0(x) +S_1(x, \bar \psi, \psi)  \\ 
		&S_0(x)= \int_0^\beta \! d\tau  \bigg[ \frac{1}{2} (\partial_\tau x) ^2 +   \frac{1}{2}\omega^2   x^2  \bigg]  \\ 
		&S_1(x, \bar \psi, \psi)  = \int_0^\beta \!d\tau  \bar \psi (\partial_\tau + {\Omega} \sigma^z+g\sqrt{\omega} x  \sigma^x)\psi.
	\end{align}	
   \end{subequations}
	We rewrite the partition function as 	
	\begin{align} \nonumber
		Z &= \int Dx \; e^{-S_0(x)} \int D\bar \psi D \psi \; e^{-S_1(x, \bar \psi, \psi)}\\
		&:=\int Dx\;  e^{-S_0(x)} Z_1 (x).
	\end{align}	
	Applying the Fourier transformation
	\begin{subequations}
	\begin{align} 
		&\psi_n = \frac{1}{\sqrt{\beta}}\int_0^\beta d\tau \psi(\tau)e^{i \nu_n\tau}, \\
		 & \bar \psi_n = \frac{1}{\sqrt{\beta}}\int_0^\beta d\tau \bar\psi(\tau)e^{-i \nu_n\tau}, \\
		& \nu_n = (2n+1)\pi /\beta,
	\end{align}	
  \end{subequations}
	we can obtain
	\begin{align} \nonumber
		Z_1 (x) = &\int D\bar \psi D\psi\exp \bigg{\{} \sum_{\nu_m,\nu_n} \bar \psi_m \big[-i\nu_n \delta_{m,n}+ {\Omega} \sigma^z \delta_{m,n} \\
		&+t x(\nu_m-\nu_n)  \sigma^x\big]\psi_n  \bigg{\}} ,
	\end{align}	
	where 
	\begin{align} 
		t = \lambda\omega\sqrt{\Omega/2\beta},    \ \ \ \ \ \ x(p_n) = \frac{1}{\sqrt{\beta}}\int_0^\beta d\tau x(\tau)e^{-i p_n\tau}.
	\end{align}
	According to the Gaussian integral, we have 
	\begin{align} \nonumber
		Z_1 (x) &= \det \big[ G_0^{-1}+t V\big] \\
		&=\det \big[ G_0^{-1}+t x(\nu_m-\nu_n)  \sigma^x \big] := \exp({-\tilde{S}}).
	\end{align}	
	where $G_0 = (-i\nu_n + {\Omega} \sigma^z)^{-1}$ is the Green function of the free spinor field,
	and the matrix $V$ satisfies $V_{m,n} = x(\nu_m-\nu_n)  \sigma^x$.
	The corrected action $\tilde{S}$ can be obtained as
	\begin{align} 
		\tilde{S} =	-\ln \det \big[ G_0^{-1}+t V\big]  = -\text{Tr} \ln \big[ G_0^{-1}+t V \big].
	\end{align}	
	Here the factor $t$ is small, so we can use the Taylor expansion
	\begin{align} \label{Sexpan}\nonumber
		\tilde{S} =&-\text{Tr} \ln \big[ G_0^{-1}+t V \big] = -\text{Tr} \ln  G_0^{-1}  -\text{Tr} \ln \big[ 1+tG_0 V \big] \\
		=&-\text{Tr} \ln  G_0^{-1}  +\sum_n \frac{t^{2n}}{2n} \text{Tr} \big[ G_0 V \big]^{2n},
	\end{align}	
	where the first term $\text{Tr} \ln  G_0^{-1} $ is a constant.

	First, we calculate the second-order contribution of Eq.~(\ref{Sexpan}), i.e.,
	\begin{align} \nonumber
		\tilde{S}_2  &= \frac{t^{2}}{2} \text{Tr} \big[ G_0 V \big]^{2}=  \frac{t^{2}}{2}\sum_{\nu_n,\nu_m} \text{Tr} [G_0(\nu_n)V_{n,m}G_0(\nu_m)V_{m,n}]\\ \nonumber
		&=   {t^{2}}\sum_{\nu_n, p_n} |x(p_n)|^2 \text{Tr}\frac{1}{-i\nu_n + {\Omega} \sigma^z}\sigma^x\frac{1}{-i\nu_n-ip_n + {\Omega} \sigma^z}\sigma^x\\
		&:=  {t^{2}}\sum_{ p_n}\pi(p_n) |x(p_n)|^2.
	\end{align}	
	Now we calculate $\pi(p_n)$ as
	\begin{align} \nonumber
		\pi(p_n) =& \sum_{\nu_n} \text{Tr}\frac{1}{-i\nu_n + {\Omega} \sigma^z}\sigma^x\frac{1}{-i\nu_n-ip_n + {\Omega} \sigma^z}\sigma^x\\ \nonumber
		=& \sum_{\nu_n} \text{Tr}\frac{1}{-i\nu_n + {\Omega} \sigma^z}\frac{1}{-i\nu_n-ip_n - {\Omega} \sigma^z}\\ \nonumber
		 =  &\sum_{\nu_n}\big(\frac{1}{-i\nu_n + {\Omega} }\frac{1}{-i\nu_n-ip_n - {\Omega} }\\ 
		&+\frac{1}{-i\nu_n - {\Omega} }\frac{1}{-i\nu_n-ip_n + {\Omega} }\big)
	\end{align}	
	Here we apply the common method of Matsubara frequency summation.
	Let 
	\begin{subequations}
	\begin{align} 
		&\pi_1(p_n) = \sum_{\nu_n}h(\nu_n)=\sum_{n_n}\frac{1}{-i\nu_n + {\Omega} }\frac{1}{-i\nu_n-ip_n - {\Omega} },\\
		&g(z) = \frac{\beta}{e^{\beta z}+1}.
	\end{align}	
   \end{subequations}
	Then we can introduce a contour integration
	\begin{align} \nonumber
		I : &= \lim_{R\rightarrow\infty}\oint\frac{dz}{2\pi i} g(z)h(-iz) \\ \nonumber
		&= \lim_{R\rightarrow\infty}\oint\frac{dz}{2\pi i} \frac{1}{-z + {\Omega} }\frac{1}{-z-ip_n - {\Omega} }\frac{\beta}{e^{\beta z}+1} \\
		&= \sum_{z_k} \text{Res}[g(z)h(-iz),z_k],
	\end{align}	
	where $\text{Res}[f(z), z_k]$ is the residue of $f(z)$ at $z_k$.
	We can find that $i\nu_n= 2i\pi n/\beta$ is the singularity of $g(z)h(-iz)$, and the corresponding residue is
	\begin{align} 
		\text{Res}[g(z)h(-iz),i\nu_n]=h(\nu_n).
	\end{align}	
	In addition to $i\nu_n$, another two singularities of $g(z)h(-iz)$ are $z_1 = \Omega$ and $z_2 = -\Omega-ip_n$, and the corresponding residues are
	\begin{subequations}
	\begin{align} 
		&\text{Res}[g(z)h(-iz),z_1]=\frac{1}{ip_n + 2{\Omega} }\frac{\beta }{e^{\beta\Omega}+1},\\
		&\text{Res}[g(z)h(-iz),z_2]=\frac{1}{ip_n + 2{\Omega} }\frac{\beta e^{\beta\Omega} }{1-e^{\beta\Omega}}
	\end{align}	
    \end{subequations}
	Therefore, we have 
	\begin{align} \nonumber
		I : = &\sum_{\nu_n}h(\nu_n)+\text{Res}[g(z)h(-iz),\Omega]\\
		&+\text{Res}[g(z)h(-iz),-\Omega-ip_n].
	\end{align}	
	Meanwhile, since we consider the  infinite radius of the contour, we have $I=0$.
	Thus, 
	\begin{align} \nonumber
		\pi_1(p_n)  = &\sum_{\nu_n}h(\nu_n)\\ \nonumber
		=&-\text{Res}[g(z)h(-iz),\Omega]-\text{Res}[g(z)h(-iz),-\Omega-ip_n]\\
		= &-\frac{\beta}{ip_n + 2{\Omega} }\frac{e^{2\beta\Omega} +1}{e^{2\beta\Omega}-1}.
	\end{align}	
	Similarly, we also have
	\begin{align} \nonumber
		\pi_2(p_n)  &= \sum_{\nu_n}\frac{1}{-i\nu_n - {\Omega} }\frac{1}{-i\nu_n-ip_n + {\Omega} }\\
		&= \frac{\beta}{ip_n - 2{\Omega} }\frac{e^{2\beta\Omega} +1}{e^{2\beta\Omega}-1}.
	\end{align}	
	Hence, we can obtain $\pi(p_n)$ as
	\begin{align} \nonumber
		\pi(p_n)  &= \pi_1(p_n)+ \pi_2(p_n) = \big(\frac{\beta}{ip_n - 2{\Omega} }-\frac{\beta}{ip_n + 2{\Omega} }\big) \frac{e^{2\beta\Omega} +1}{e^{2\beta\Omega}-1}\\
		&= -\frac{4\beta\Omega}{p_n^2 + 4{\Omega}^2 }\coth(\beta \Omega).
	\end{align}	
	Since $\beta\rightarrow\infty$, we have $\coth(\beta \Omega)=1$.
	In addition, we mainly concern the infrared limit, i.e., $p_n\ll \Omega$, thus $\pi(p_n)$ can be approximated as 
	\begin{align} 
		\pi_2(p_n)  \approx \frac{\beta p_n^2}{ 4{\Omega}^3 }-\frac{\beta }{ {\Omega} }.
	\end{align}	
	Therefore, the second-order contribution of Eq.~(\ref{Sexpan}) is
	\begin{align} \nonumber
		\tilde{S}_2  &=  {t^{2}}\sum_{ p_n} \big(\frac{\beta p_n^2}{ 4{\Omega}^3 }-\frac{\beta }{ {\Omega} }\big) |x(p_n)|^2\\
		&=\frac{1}{2}\sum_{ p_n} \big(\frac{\lambda^2\omega^2 p_n^2}{ 4{\Omega}^2 }-\lambda^2\omega^2  \big) |x(p_n)|^2.
	\end{align}

	Next, we calculate the 4th-order contribution of Eq.~(\ref{Sexpan})
	\begin{widetext}
	\begin{align} \label{s4}\nonumber
		\tilde{S}_4  =& \frac{t^{4}}{4} \text{Tr} \big[ G_0 V \big]^{4}
		=  \frac{t^{4}}{4}\sum_{\nu_n,\nu_m,\nu_\ell,\nu_k} \text{Tr} [G_0(\nu_n)V_{n,m}G_0(\nu_m)V_{m,\ell}G_0(\nu_\ell)V_{\ell,k}G_0(\nu_k)V_{k,n}]\\ \nonumber
		=  &\frac{t^{4}}{4} \sum_{\nu_n, p_n,q_n,r_n} x(p_n) x(q_n)x(r_n)x(-p_n-q_n-r_n)\\
		&\text{Tr}\frac{1}{-i\nu_n + {\Omega} \sigma^z}\sigma^x\frac{1}{-i\nu_n-ip_n + {\Omega} \sigma^z}\sigma^x\frac{1}{-i\nu_n-ip_n-iq_n + {\Omega} \sigma^z}\sigma^x\frac{1}{-i\nu_n-ip_n-iq_n-ir_n + {\Omega} \sigma^z}\sigma^x.
	\end{align}	
   \end{widetext}
	Here, $\tilde{S}_4 $ can also be obtained by Matsubara frequency summation.
	However, we only consider the long-wavelength limit rather than the explicit form of $\tilde{S}_4 $.
	Thus, we can neglect the microscopic details, where $\tilde{S}_4 $ can be described by a simple form
	\begin{align} 
		\tilde{S}_4  = a_4  \omega^4 \int_0^\beta d\tau   x^4,
	\end{align}	
	where $a_4$ is a function of $\lambda$ and $\Omega$.
	Similarly, we can also give estimations of higher-order contributions as
	\begin{align} 
		\tilde{S}_{2n}  = a_{2n}  \omega^{2n} \int_0^\beta d\tau   x^{2n}.
	\end{align}	
	Therefore, the effective action of the harmonic  oscillation can be obtained as 
	\begin{align} \nonumber
		S_{\text{eff}} = &S_0 + \tilde{S_2}+ \tilde{S_4} + \tilde{S_6} + ... \\ \nonumber
		=& \int_0^\beta d\tau  \bigg[ \frac{1}{2} (1+ \frac{\lambda^2\omega^2}{4\Omega^2})(\partial_\tau x) ^2 \\
		&+\frac{1}{2}(1-\lambda^2)  \omega^2x^2 + a_4 \omega^4 x^4+  a_6 \omega^6 x^6+...\bigg].
	\end{align}

	\section{Effective action of the Dicke model}\label{app2}
	In this section, we consider the Dicke model, which describes light interacting with a large ensemble of two-level atoms.
	The Hamiltonian reads
	\begin{align} \label{Hdicke}
		\hat H_{\text{Dicke}} = \omega \hat a^\dag \hat a + {\Omega}\sum_{j=1}^N\hat \sigma^z_j+\frac{2J}{\sqrt{N}}(\hat a^\dagger+\hat a)\sum_{j=1}^N\hat \sigma^x_j.
	\end{align}	
	where $\hat \sigma^\alpha_j (\alpha=x,y,z)$ are the Pauli matrices describing the $j$-th two-level atom,
	and $N$ is the number of atoms.
	There is also a parity symmetry $[\hat P, \hat H_{\text{Dicke}}]=0$ with $\hat P = (-1)^{\hat n}\prod_{j=1}^N\hat \sigma_j^z$.
	In the thermodynamic limit $N\rightarrow \infty$, 
	there also exists a second-order QPT in this zero-dimensional system from the normal phase to the superradiant phase when increasing $\lambda$.
	The critical point is exact at $\lambda=\lambda_c=\sqrt{\omega\Omega}/2$, and this QPT can also be described by the spontaneous breaking of the parity symmetry.
	
	Here, to obtain the effective action of the Dicke model, we need to apply the Holstein-Primakoff transformation.
	We first introduce the angular momentum representation 
	\begin{align} 
		\hat S^\alpha : = \frac{1}{2}\sum_{j=1}^N\hat \sigma^\alpha_j,
	\end{align}	
	where $\hat S^\alpha$ is the spin-$N/2$ angular-momentum operator.
	According to the Holstein-Primakoff transformation, we have
	\begin{subequations}
	\begin{align} 
		&	\hat S^+  = (\sqrt{N-\hat b^\dagger\hat b})\hat b ,\\ 
		&	\hat S^-  = \hat b^\dagger(\sqrt{N-\hat b^\dagger\hat b}),\\
		&	\hat S^z  = N/2-\hat b^\dagger\hat b,
	\end{align}	
   \end{subequations}
	where $\hat b^\dagger (\hat b)$ is the bosonic creation (annihilation) operator.
	Thus, the Hamiltonian in Eq.~(\ref{Hdicke}) can be written as
	\begin{align} \label{Hdicke2} \nonumber
		\hat H_{\text{Dicke}} =& \omega \hat a^\dag \hat a + 2{\Omega}\hat b^\dagger\hat b\\
		&+\frac{4J}{\sqrt{N}}(\hat a^\dagger+\hat a)\big(\sqrt{N-\hat b^\dagger\hat b}\ \hat b+ \hat b^\dagger\sqrt{N-\hat b^\dagger\hat b}\big).
	\end{align}	
	Here, we can use the Taylor expansion to expand the square root term as
	\begin{align} 
		\sqrt{N-\hat b^\dagger\hat b} = \sqrt{N}\big(1-\frac{\hat b^\dagger\hat b}{2N} +\frac{(\hat b^\dagger\hat b)^2}{4N^2} + ...\big).
	\end{align}	
	Thus we have 
	\begin{align}\nonumber
		&\hat H_{\text{Dicke}} = \omega \hat a^\dag \hat a + 2{\Omega}\hat b^\dagger\hat b+4J(\hat a^\dagger+\hat a)(\hat b^\dagger+\hat b)\\
		&+ 4J(\hat a^\dagger+\hat a)\big(-\frac{\hat b^\dagger\hat b\hat b+\hat b^\dagger\hat b^\dagger\hat b}{2N}+\frac{(\hat b^\dagger\hat b)^2\hat b+\hat b^\dagger(\hat b^\dagger\hat b)^2}{4N^2}+....\big).
	\end{align}	
	Therefore, according to the interaction between the bosons $\hat a$ and $\hat b$, we can write a general effective imaginary-time action of the Dicke model as
	\begin{align}\nonumber
		S_{\text{eff}} (x)=& \int_0^\beta d\tau  \big[\frac{1}{2} (\partial_\tau \tilde{x}) ^2 + 
		\frac{2\omega}{\Omega}(\lambda_c^2-\lambda^2) \tilde{x}^2 \\
		&+ \frac{\alpha_4}{N}  \tilde{x}^4+   \frac{\alpha_6}{N^2} \tilde{x}^6+.... \big],
	\end{align}	
	where $\tilde{x}$ is also a real scalar field, and  $\alpha_{2n}$ is a finite factor with its explicit form unimportant.
	We note that the $\tilde{x}$ here is not the coordinate of the bare oscillator $\hat a$, while it is a linear combination of $\hat a$ and $\hat b$.
	\end{appendix}

\end{document}